\documentclass[twocolumn,amsmath,amssymb,prapplied,longbibliography]{revtex4-2}

\usepackage{graphicx}
\usepackage{dcolumn}
\usepackage{bm}
\usepackage{color}
\usepackage{notes2bib}

\begin{document}



\title{Nuclear spin polarization in silicon carbide at room temperature in the Earth's magnetic field}

\author{A.~N.~Anisimov$^{1}$}
\email[E-mail:~]{a.anisimov@hzdr.de}
\author{A.~V.~Poshakinskiy$^{2}$}
\author{G.~V.~Astakhov$^{1}$}
\email[E-mail:~]{g.astakhov@hzdr.de}

\affiliation{$^1$Helmholtz-Zentrum Dresden-Rossendorf, Institute of Ion Beam Physics and Materials Research, 01328 Dresden, Germany  \\
$^2$ICFO-Institut de Ciencies Fotoniques, The Barcelona Institute of Science and Technology, 08860 Castelldefels, Barcelona, Spain
 }

\begin{abstract}
Coupled electron-nuclear spins represent a promising quantum system, where the optically induced electron spin polarization can be dynamically transferred to nuclear spins via the hyperfine interaction. Most experiments on dynamic nuclear polarization (DNP) are performed at cryogenic temperatures  and/or in moderate external magnetic fields, the latter approach being very sensitive to the magnetic field orientation.  Here, we demonstrate that the  $\mathrm{^{29}}$Si nuclear spins 
in SiC 
can be efficiently polarized at room temperature even in the Earth's magnetic field. We exploit the asymmetric splitting of the optically detected magnetic resonance (ODMR) lines inherent to half-integer $S = 3/2$ electron spins,  
such that  the certain transitions involving $\mathrm{^{29}}$Si nuclei can be clearly separated and selectively addressed using radiofrequency (RF) fields.  As a model system, we use the  V3 silicon vacancy ($\mathrm{V_{Si}}$) in 6H-SiC, which has the zero-filed splitting parameter comparable with the hyperfine interaction constant. Our theoretical model considers  DNP under optical excitation in combination with RF driving and agrees very well with the experimental data. In the case of high-fidelity electron spin polarization, the proposed DNP protocol leads to ultra-deep optical cooling of nuclear spins with an effective temperature of about $50 \, \mathrm{n K}$. These results provide a straightforward approach for controlling the nuclear spin under ambient conditions, representing an important step toward realizing nuclear hyperpolarization for magnetic resonance imaging and long nuclear spin memory for quantum logic gates.

\end{abstract}

\date{\today}

\maketitle

\section{Introduction}

Nuclear spins are considered as a very promising quantum system for quantum information processing and sensing. Due to their very long spin coherence time, they are a key component of spin-based quantum registers \cite{10.1126/science.1139831, 10.1103/PhysRevX.9.031045, 10.1038/s41563-020-00802-6, 10.1038/s41563-021-01148-3}. Furthermore, nuclear spin (hyper)polarization can enhance the signal intensities in magnetic resonance imaging (MRI) by several orders of magnitude allowing detection of small chemical shifts and analysis of single cells \cite{10.1021/acs.chemrev.2c00534, 10.1016/j.copbio.2023.102975}. Due to the small value of the nuclear magneton, an effective way to initialize nuclear spins is first to polarize  electron spins using optical excitation  and then to transfer this polarization to the nuclear spin via the hyperfine interaction. Indeed, such dynamic nuclear polarization (DNP) has been demonstrated in a variety of materials, including GaAs \cite{10.1038/s42005-021-00681-6}, diamond \cite{10.1126/science.1131871, 10.1103/physrevlett.102.057403} and silicon carbide (SiC) \cite{10.1103/physrevlett.114.247603, 10.1103/physrevlett.132.090601}. 

The nitrogen-vacancy (NV) defect in diamond is one of the most studied solid-state system, which demonstrates quantum properties even at room temperature. Using selective microwave excitation and controlled Larmor precession in an external magnetic field, a high degree of  nuclear spin polarization coupled to single NV defects has been reported \cite{10.1126/science.1131871, 10.1126/science.1139831}. This approach provides only random initialization of the nuclear spin states, which is not applicable for a large number of qubits or for the hyperpolarization-enhanced MRI. This problem can be overcome using the DNP scheme  based on optical pumping and  level anticrossing (LAC) of the NV electron spin states in a certain magnetic field \cite{10.1103/physrevlett.102.057403}. This approach is extremely simple, but it relies on a very precise orientation of the external magnetic field with respect to the NV defect axis.  Alternatively, an advanced DNP protocol with multiple microwave fields in adiabatically rotating magnetic fields has been theoretically proposed to achieve nuclear hyperpolarization even in a randomly oriented ensemble of nanodiamonds \cite{10.1103/physrevb.92.184420}.  

The lack of wafer-scale fabrication for diamond limits its industrial application in the field of quantum technologies. The divacancies (VV) and silicon vacancies ($\mathrm{V_{Si}}$) in SiC have attracted much interest as an alternative platform, as they possess appealing quantum properties  in a technologically mature material \cite{10.1038/nature10562, 10.1103/physrevlett.109.226402}. SiC crystals with natural isotope abundance contain 4.7\% of  $\mathrm{^{29}}$Si and  1.1\% of  $\mathrm{^{13}}$C, both isotopes have non-zero nuclear spin $I = 1/2$. The VV center in SiC has electron spin $S=1$, similar to the NV defect in diamond. Correspondingly,  the $\mathrm{^{29}}$Si DNP  has been demonstrated under optical pumping using the ground-state and excited-state level anticrossing (GSLAC and ESLAC), which is strongest for the magnetic field $B = 33 \, \mathrm{mT}$ and drops rapidly if the magnetic field is misaligned from the VV defect axis by several degrees \cite{10.1103/physrevlett.114.247603, 10.1103/physrevb.92.115206}. Following this approach, room-temperature entanglement between the VV electron spin and coupled  $\mathrm{^{29}}$Si spin in a macroscopic ensemble has been reported \cite{10.1126/sciadv.1501015}.

\begin{figure*}[t]
\includegraphics[width=.95\textwidth]{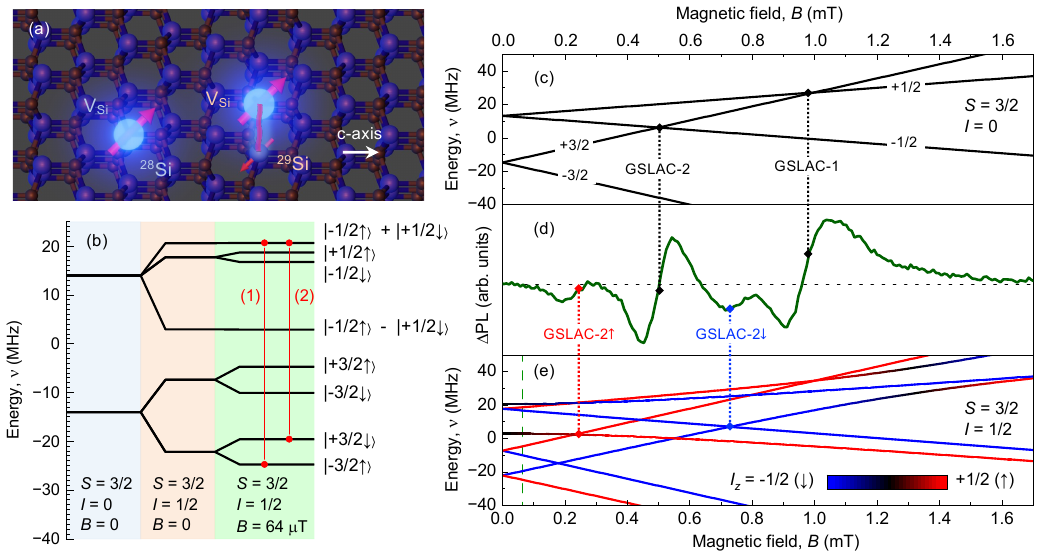}
\caption{Electron-nuclear spin coupling in the V3 $\mathrm{V_{Si}}$ center in 6H-SiC. (a)
Schematic presentation of the optically-active silicon vacancy $\mathrm{V_{Si}}$ with electron spin $S = 3/2$. Left: 12 next nearest neighbor $^{28}$Si  nuclei with spin $I = 0$.  Right: 11 next nearest neighbor $^{28}$Si  nuclei with spin $I = 0$ and one nearest neighbor $^{29}$Si  nucleus with spin $I = 1.2$.  (b) Energy levels of the V3 $\mathrm{V_{Si}}$ center in the ground state with $I = 0$ and  $I = 1/2$ in zero magnetic field and nonzero magnetic field $B = 64 \, \mathrm{\mu} T$.  The nuclear spin polarization is induced and detected using the spin transitions labeled as (1) and (2). (c) Zeeman shift of the V3 $\mathrm{V_{Si}}$ spin sublevels with $S = 3/2$ and $I = 0$ in external magnetic fields applied parallel to the c-axis. The spin states with $S_Z = \pm 3/2, \pm 1/2$ are labeled. The vertical dotted lines indicate GSLAC-2 and GSLAC-1. (d)   PL variation of the V3 $\mathrm{V_{Si}}$ in external magnetic fields. For the lock-in detection, a weak oscillating magnetic field is additionally applied.  (e)  Zeeman shift of the V3 $\mathrm{V_{Si}}$ spin sublevels with $S = 3/2$ and $I = 1/2$ in external magnetic fields applied parallel to the c-axis.  The color codes the nuclei spin projection $I_z$ for each spin state. The vertical dotted lines indicate GSLAC-2$\uparrow$ and GSLAC-2$\downarrow$.  The vertical dashed line at  $B = 64 \, \mathrm{\mu} T$ indicates the magnetic field used for the nuclear spin polarization. }   \label{fig1}
\end{figure*}

In contrast to the NV and VV defects, the optically active $\mathrm{V_{Si}}$ defects possess half-integer spin $S = 3/2$ and have some features not inherent to the systems with integer spin \cite{10.1103/PhysRevB.71.201312, 10.1038/nphys2826, 10.1103/physrevx.6.031014, 10.1038/srep33301, 10.1103/physrevb.95.081405, 10.1038/s41467-019-09873-9, 10.1038/s41563-021-01148-3, 10.1038/s41467-019-09429-x, 10.1103/physrevlett.125.107702, 10.1103/physrevb.101.144109, 10.1038/s41534-020-00310-0, 10.1126/sciadv.abj5030, 10.1038/s41928-023-01029-4}. This includes GSLAC- and ESLAC-based sensing with exceptional sensitivity \cite{10.1103/physrevx.6.031014, 10.1038/srep33301, 10.1103/physrevb.95.081405},  spectrally stable and reconfigurable single photon emitters coupled to electron spins  \cite{10.1038/s41467-019-09873-9, 10.1038/s41534-020-00310-0, 10.1038/s41563-021-01148-3} and microwave-free acoustic control of spins \cite{10.1103/physrevlett.125.107702, 10.1126/sciadv.abj5030, 10.1038/s41928-023-01029-4}. The DNP of $\mathrm{^{29}}$Si spins coupled to the $\mathrm{V_{Si}}$ spins has been realized in the vicinity of the GSLAC at low temperature only \cite{10.1103/physrevlett.132.090601}. In addition, the application of moderate magnetic fields suppresses the heteronuclear spin pair flip-flop, i.e., between  the $\mathrm{^{29}}$Si  and $\mathrm{^{13}}$C nuclear spins \cite{10.1103/physrevb.90.241203, 10.1103/physrevb.95.161201}, which spatially localize the DNP and may prevent generation of nuclear hyperpolarization in an ensemble under optical pumping.  

In this work, we report an efficient nuclear spin polarization in 6H-SiC at room temperature in the Earth's magnetic field. The DNP is realized due to the non-symmetric hyperfine splitting of the $\mathrm{V_{Si}}$ spin states coupled to a $\mathrm{^{29}}$Si nuclear spin $I = 1/2$, which is unique for spin-3/2 centers. The singlet states in this electron-nuclear spin system have zero nuclear spin, in contrast to the doublet states, which all have non-zero nuclear spin. The radiofrequency (RF) induced transition between any of the singlet and doublet states, followed by optical repumping, results in nuclear spin polarization. The Earth's magnetic field is only needed to remove the state degeneracy and spectrally select a particular nuclear spin orientation. Our approach is also suitable for the generation of DNP in SiC nanoparticles \cite{10.1063/1.4904807}, which are proposed as hyperpolarized MRI agents \cite{10.1021/acsami.1c07156} and the DNP can be more easily implemented compared to nanodiamonds due to the much weaker magnetic fields required \cite{10.1103/physrevb.92.184420}.

\begin{table*}[t]
\setlength{\tabcolsep}{8pt}
\setlength{\extrarowheight}{4pt}
    \centering
\newcommand{\minus}{\raisebox{.1\height}{\scalebox{.8}{--}}}
\newcommand{\plus}{\raisebox{.2\height}{\scalebox{.66}{+}}}
\begin{tabular}{c|c |c c c c c}
       $|S_z, I_z\rangle$  &    $S_zI_z$   &  $|\plus\frac12,\downarrow \rangle + |\minus\frac12,\uparrow \rangle$ & $|\plus\frac12,\uparrow \rangle$; $|\minus\frac12,\downarrow \rangle$ & $|\plus\frac12,\downarrow \rangle - |\minus\frac12,\uparrow \rangle$ &  $|\plus\frac32,\uparrow \rangle$; $|\minus\frac32,\downarrow \rangle$ & $|\plus\frac32,\downarrow \rangle$; $|\minus\frac32,\uparrow \rangle$ \\[.1cm]
       \hline
$|\plus\frac12,\downarrow \rangle + |\minus\frac12,\uparrow \rangle$ & $\minus\frac14$&  -- & $|A|/2$  & $2|A|$ & $\times$ & ${|2D-3A/2|}$\\
$|\plus\frac12,\uparrow \rangle$; $|\minus\frac12,\downarrow \rangle$& $\plus\frac14$ & $|B_\pm|^2$ &-- & $3|A|/2$ & ${|2D+A/2|}$ & $|2D-A|$\\
$|\plus\frac12,\downarrow \rangle - |\minus\frac12,\uparrow \rangle$ & $\minus\frac14$ & $\frac14 |B_z|^2$ & $|B_\mp|^2$ & -- & $\times$ & ${|2D+A/2|}$\\
$|\plus\frac32,\uparrow \rangle$; $|\minus\frac32,\downarrow \rangle$& $\plus\frac34$ & $\times$ & $\frac32 |B_\pm|^2$ & $\times$ & -- & $3|A|/2$ \\
$|\plus\frac32,\downarrow \rangle$; $|\minus\frac32,\uparrow \rangle$& $\minus\frac34$ & $\frac34 |B_\pm|^2$ & $\frac3{4} \big(\frac{A}{2D-A}\big)^2 |B_z|^2$ & $\frac34 |B_\pm|^2$ & $\frac9{8} \big(\frac{A}{2D-A}\big)^2 |B_\mp|^2$ & --\\
    \end{tabular}
    \caption{The dipole-allowed transitions between the fine structure components in zero static magnetic field. 
    The upper part of the table indicates the transition energy. The lower part of the table shows squared absolute value of the transition matrix element. The calculation is performed after Hamiltonian~\eqref{eq:H} in the limit of $A_\parallel = A_\perp \equiv A \ll |2D|$. $B_z$ and $B_\pm = (B_x \mp i B_y)/\sqrt2$ are the RF field components.}
    \label{tab:hyperfine}
\end{table*}

\section{Hyperfine structure of the silicon vacancy}

As a model system, we use the V3  $\mathrm{V_{Si}}$ defect in 6H-SiC \cite{10.1103/physrevb.83.125203,10.1038/s41534-022-00534-2}, which  is schematically depicted in Fig.~\ref{fig1}(a). It possesses spin $S = 3/2$ \cite{10.1038/nphys2826} and is surrounded by  4 C atoms as  the nearest neighbors and 12 Si atoms as the next nearest neighbors (NNN). With the natural isotope abundance of 4.7\% for $\mathrm{^{29}}$Si, 
the probability that a single $\mathrm{V_{Si}}$ defect has zero nuclear spin among the NNN (i.e., $\mathrm{^{28}}$Si or $\mathrm{^{30}}$Si isotopes) or coupled to a single nuclear $\mathrm{^{29}}$Si spin with $I = 1/2$ is equal 56\% or 33\%, respectively.

The fine and hyperfine structure of the $\mathrm{V_{Si}}$ coupled to a $\mathrm{^{29}}$Si nuclear spin is described by the effective Hamiltonian
\begin{align}\label{eq:H}
H = &D \left[S_z^2 - \frac{S(S+1)}{3}\right] + g\mu_B \bm B \cdot \bm S \nonumber\\
&+ A_\parallel S_z I_z + A_\perp (S_xI_x+S_yI_y) ,
\end{align}
where $\bm S$ and $\bm I$ are the operators of the electron and nuclear spin, $D/h=-14\,$MHz is the zero-field splitting parameter \cite{10.1103/physrevb.83.125203, 10.1103/PhysRevB.66.155214}, $A_\parallel$ and $A_\perp$ are the hyperfine interaction constants, that are nearly spherically-symmteric $A_\parallel = A_\perp \equiv A$ with $A/h =  8.8 \,$MHz~\cite{10.1103/PhysRevB.66.155214,10.1103/PhysRevB.104.125205}, $g=2$ is the electron $g$-factor, $\mu_B$ is the Bohr magneton, and $\bm B$ is the external magnetic field. The latter can consist of a static field $B_z$ applied along the $c$-axis, and an RF field $\bm B_1$ that induces spin transitions. The effect of magnetic field on the nuclear spin is neglected due to much smaller value of the nuclear magneton compared to $\mu_B$.

The spin level structure calculated after Eq.~\eqref{eq:H} is presented in Fig.~\ref{fig1}(b). The first term in Eq.~\eqref{eq:H} leads to the splitting of the electronic states with $S_z = \pm 1/2$ and $\pm 3/2$ by the energy $2D$, i.e., the first column of Fig.~\ref{fig1}(b). This fine structure is observed for the $\mathrm{V_{Si}}$ defects with zero nuclear spin among the 12 NNN. If there is one $\mathrm{^{29}}$Si among the 12 NNN,  the effect of hyperfine interaction is shown in the central column of Fig.~\ref{fig1}(b). The longitudinal component of the hyperfine interaction $A_\parallel$ splits the energy levels into 4 doublets,  charecterized by $S_zI_z = \pm 1/4$ and $\pm 3/4$ (table~\ref{tab:hyperfine}). 

The doublet with $S_zI_z = - 1/4$ (table~\ref{tab:hyperfine}) has the states that differ in electron and nuclear spin projection by $\pm 1$ and can be further split by the transverse component of the hyperfine interaction $A_\perp$ into a pair of singlets $|s\rangle, |s'\rangle = (|+1/2,\downarrow \rangle \pm |-1/2,\uparrow \rangle)/\sqrt{2}$. The appearance of these mixed singlets states is a unique feature of the hyperfine structure of spin-3/2 centers (or other spin centers with higher half-integer spin $S \geq 3/2$ \cite{10.1103/PhysRevResearch.4.033107}). For the spin centers with integer spin, for instance the NV defect in diamond or divacancies in SiC, the transverse component of the hyperfine interaction $A_\perp$ has almost no effect on the fine structure, which consists of $2S+1$ doublets that cannot be mixed by $A_\perp$.  

The third column of Figure~\ref{fig1}(b) shows the effect of a weak magnetic field $B \ll A_\perp/\mu_B$, 
which splits the doublets  but does no affect the mixed singlet states. In contrast, in a strong magnetic field the effect of state mixing by transverse hyperfine interaction would be suppressed. 

The existence of the mixed singlet states for our system with $S=3/2$ and $I =1/2$ in small magnetic fields leads to an important consequence, that would be absent for centers with integer spin and for any centers in large magnetic fields. Namely, the energies of the transitions that involve singlet states are distributed in a non-symmetric manner with respect to the central energy $|2D| = 28 \, \mathrm{MHz}$. There is a transition at the energy of $ |2D+A/2| = 23.6 \, \mathrm{MHz}$ between the doublets with $S_zI_z = +1/4$ and $+3/4$, as also summarized in table~\ref{tab:hyperfine}. There is also a transition at the same energy of $ |2D+A/2| = 23.6 \, \mathrm{MHz}$ between the doublet with $S_zI_z = -3/4$ and one of the singlets $S_zI_z = -1/4$. And the transition between the doublet with $S_zI_z = -3/4$ and the second singlet $S_zI_z = -1/4$ has different transition energy of $ |2D - 3A/2| = 41.2 \, \mathrm{MHz}$. In an external magnetic field, this doubly degenerate transition is split in two transitions depending on the nuclear spin orientation, which are labeled as (1) and (2) in Fig.~\ref{fig1}(b). 

In order to experimentally verify the hyperfine structure, we start from the RF-free photolumenescence (PL) spectroscopy of the V3 $\mathrm{V_{Si}}$ defects in external magnetic fields \cite{10.1007/s00723-017-0938-1}. Figure~\ref{fig1}(c) shows the Zeeman shift of the $S_z = \pm 3/2, \pm 1/2$ levels for $I = 0$. There are two characteristic magnetic fields of $1 \, \mathrm{mT}$ and $0.5 \, \mathrm{mT}$, corresponding to the GSLAC-1 between the $S_z = + 3/2$ and $S_z = + 1/2$ states and the GSLAC-2 between the $S_z = + 3/2$ and $S_z = - 1/2$ states \cite{10.1103/physrevx.6.031014}. These magnetic fields are clearly imprinted in the PL intensity variation $\mathrm{\Delta PL}$ of Fig.~\ref{fig1}(d).  The level structure in case of $I = 1/2$ has much complex behavior in external magnetic fields, as presented in Fig.~\ref{fig1}(e). However, one can clearly identify the GSLAC-2$\uparrow$ at $B = 0.25 \,  \mathrm{mT}$, which arises from the the nuclear spin state $I_z = +1/2$. Correspondingly, there is also the GSLAC-2$\downarrow$ at $B = 0.75 \,  \mathrm{mT}$ for the nuclear spin state $I_z = -1/2$. These nuclear spin satellites are well observed in the $\mathrm{\Delta PL}$ signal plotted against external magnetic field shown in Fig.~\ref{fig1}(d). Similar behavior has been recently observed for the V2 $\mathrm{V_{Si}}$ defect in 4H-SiC \cite{10.1134/S0021364023603834} and proposed for an optically pumped self-calibrated magnetometry using the V3 $\mathrm{V_{Si}}$ defect in 6H-SiC \cite{10.1109/jsen.2024.3391191}.  

\begin{figure}[t]
\includegraphics[width=.45\textwidth]{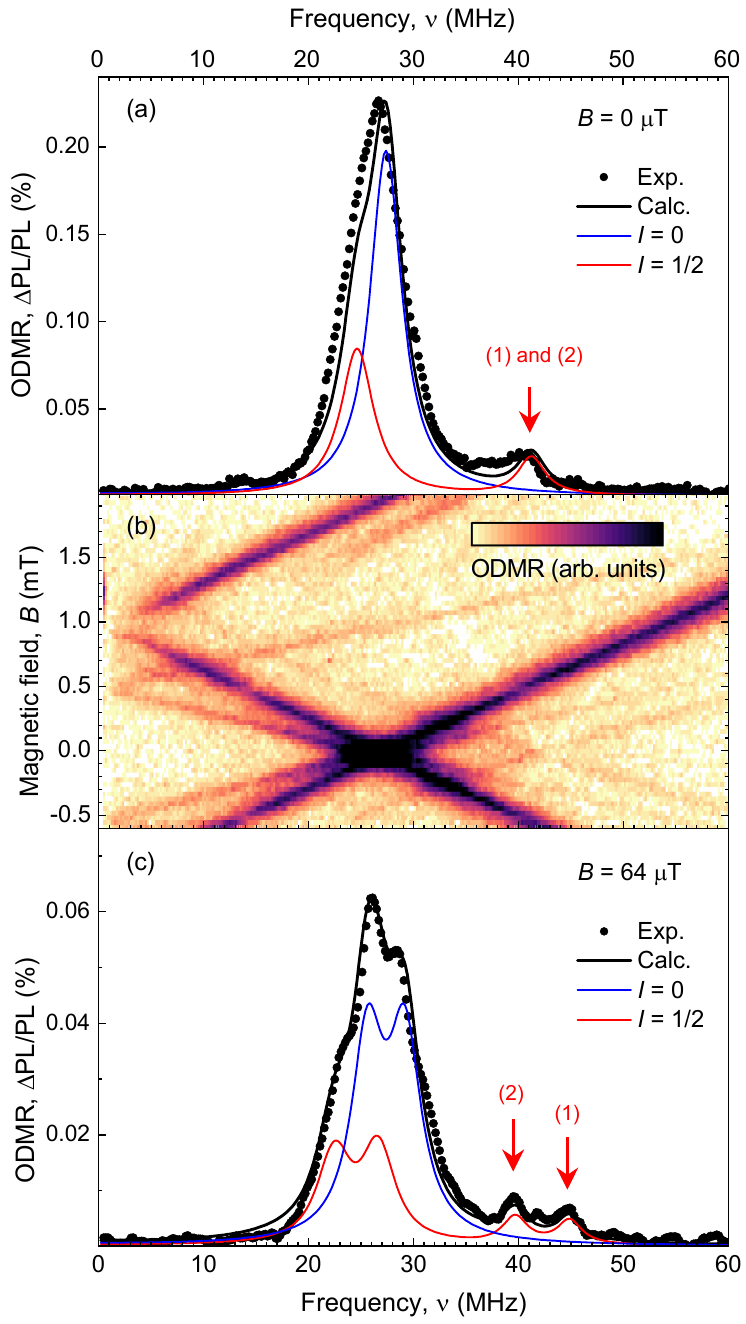}
\caption{Room-temperature ODMR spectra of the V3 $\mathrm{V_{Si}}$ center in 6H-SiC. (a) ODMR spectrum in zero magnetic field. The black line is a fit to spin Hamiltonian of Eq.~(\ref{eq:H}). The blue curve corresponds to zero $^{28}$Si  nuclear spin $I = 0$ and the red curve corresponds to nonzero $^{29}$Si nuclear spin $I = 1/2$.  The black curve is a cumulative fit. (b) Color plot of the evolution of the ODMR spectrum in magnetic fields from $-0.6 \, \mathrm{mT}$ to $2.0 \, \mathrm{mT}$. (c) The same as (a) but in an external magnetic field $B = 64 \, \mathrm{\mu T}$. The vertical arrow indicates transitions at (1) and  (2) between the spin states with $I = 1/2$ shown in  Fig.~\ref{fig1}(b). } \label{fig2}
\end{figure}

Next, we measure optically detected magnetic resonance (ODMR) spectra of the V3 $\mathrm{V_{Si}}$ center in 6H-SiC (Fig.~\ref{fig2}).  Figure~\ref{fig2}(a) shows an ODMR spectrum in zero magnetic field, which is  dominated by the peak at  $\nu_1 = 28 \, \mathrm{MHz}$ with an additional peak at around $\nu_1 = 41 \, \mathrm{MHz}$. To ensure that $B= 0 \, \mathrm{\mu T} $ and the Earth's magnetic field is compensated, we us a coil to generate an additional magnetic field and scan it from negative to positive values. A corresponding 2D map is presented in Fig.~\ref{fig2}(b).  The RF powers is selected such that to maximaize the ODMR signal at $41 \, \mathrm{MHz}$, corresponding to the degenerate transitions (1) and (2), but to suppress the multi-photon transitions and avoid significant broadening of the ODMR lines \cite{10.1103/PhysRevResearch.4.023022} (details are presented in the Supplemental materials). 

To calculate the ODMR spectrum of Fig.~\ref{fig2}(a), we use the transition energies and matrix elements from table~\ref{tab:hyperfine}. The calculation shown by the black solid line is in a very good agreement with our experimental data. The contributions to the ODMR spectrum  from the $\mathrm{V_{Si}}$ center with $I = 0$ and $I = 1/2$ are shown by the blue and red curves, respectively. Their relative contribution  $\approx 0.56/0.33$  is determined by the ratio of the probabilities to find no  $\mathrm{^{29}}$Si spins and one such spin among the 12 NNN. The only fitting parameter is the broadening of the ODMR lines. 

After turning off the current in the coil, we measure the ODMR spectrum again as shown in  Fig.~\ref{fig2}(c). From the fit of the experimental data (the solid lines), we determine the environment magnetic field $B= 64 \, \mathrm{\mu T} $.  It is contributed by the Earth's magnetic field and the remanent field of the experimental setup.  The splitting between transitions (1) and (2) in this field given by $\Delta E = 3 \mu_B B$ is clearly seen.

\section{Dynamic nuclear polarization in the Earth's magnetic field}

\begin{figure*}[t]
\includegraphics[width=.97\textwidth]{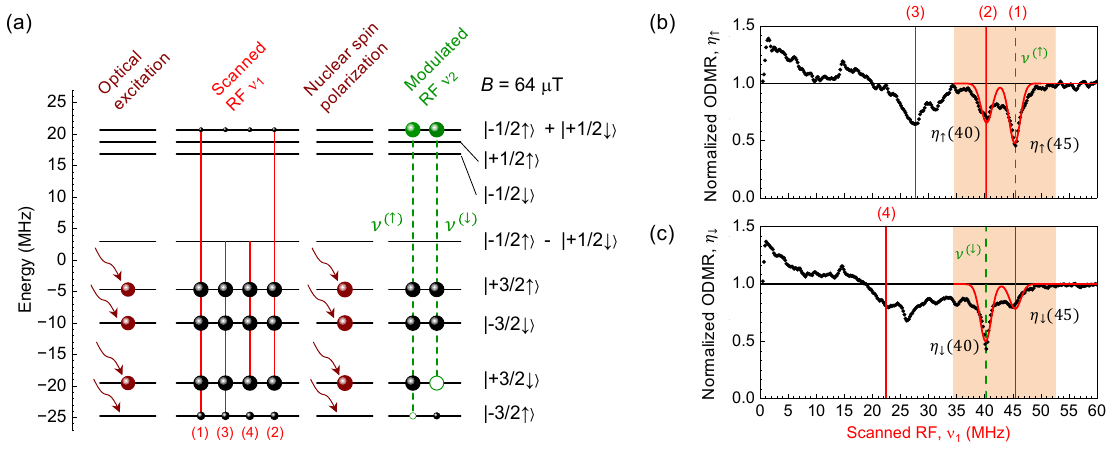}
\caption{Optically induced $^{29}$Si nuclear spin polarization at room temperature.  (a) Protocol for nuclear spin polarization in a weak magnetic field $B = 64 \, \mathrm{\mu T}$. Optical pumping of the V3 $\mathrm{V_{Si}}$ center leads to the preferential population of the $| \pm 3/2 \rangle$ states. The scanned RF $\nu_1$ induces transition between the $| \pm 3/2 \rangle$ spin states with non-zero nuclear spin to the singlet states  $| - 1/2 \uparrow \rangle \pm | + 1/2 \downarrow \rangle$ with vanishing average nuclear spin. Optical repumping results in a $^{29}$Si nuclei spin polarization, which is probed by a second RF $\nu_2$. It is tuned to certain spin transitions ($\nu^{(\uparrow)}$ or $\nu^{(\downarrow)}$) and modulated for lock-in detection. (b) ODMR spectrum as a function of the scanned RF $\nu_1$  modulated at a fixed RF $\nu_2 = \nu^{(\uparrow)}$. A Gaussian fit of two dips at  $\nu_1 = 40 \, \mathrm{MHz}$ and $\nu_1 = 45 \, \mathrm{MHz}$ is shown by the solid line. The vertical lines indicate the allowed spin transitions, labeled in (a). The vertical dashed line represents the modulated frequency $\nu_2$. (c) The same as (b) but at a modulation RF $\nu_2 = \nu^{(\downarrow)}$. The shaded area in (b) and (c) indicates the RF spectrum range used for the generation and detection of DNP. } \label{fig3}
\end{figure*}

Now, we describe the DNP protocol in the Earth's magnetic field, which ranges between approximately $25 \, \mathrm{\mu T}$ and $65 \, \mathrm{\mu T}$ \cite{10.1111/j.1365-246X.2010.04804.x}. Figure~\ref{fig3}(a) shows the energy diagram of the spin levels in the magnetic field $B= 64 \, \mathrm{\mu T}$, corresponding to our experimental conditions. Optical excitation followed by spin-dependent intersystem crossing leads to the preferential population of the $| \pm 3/2 \rangle$ states. This process can be described in terms of electron spin quadrupole polarization \cite{10.1002/pssb.201700258,10.1038/s41467-019-09429-x}
\begin{align}
\label{eq:quadrupole}
    P_Q = P_{3/2}-P_{1/2}
\end{align}
where $P_{3/2}$ and $P_{1/2}$ are the probabilities for the $\mathrm{V_{Si}}$ center to be in the state with the electron spin $|S_z| = 3/2$ and $|S_z| = 1/2$ respectively.

The singlet state $|s\rangle = |+1/2,\downarrow \rangle + |-1/2,\uparrow \rangle$ (the normalization coefficient is omitted for simplicity) has vanishing average nuclear spin, which is in contrast to the doublet states that all have $I_z = \pm 1/2$.
In particular, consider the energetically lowest $|-3/2,\uparrow \rangle$ and $|+3/2,\downarrow \rangle$ states. The RF-induced transitions between these states and the singlet state at frequencies $\nu^{(\uparrow)}= 45 \, \mathrm{MHz}$ and $\nu^{(\downarrow)}= 40 \, \mathrm{MHz}$ are accompanied with the change of nuclear spin, see the transitions marked as (1) and (2) in Fig.~\ref{fig1}(b). Therefore, they lead to the nuclear spin polarization. 

In our experiments, we excite the system with two frequencies simultaneously. The first RF $\nu_1$ induces DNP and the second RF $\nu_2$ is used to measure it.
To suppress the inhomogeneous broadening and increase the signal-to-noise ratio, we use the spectral hole burning technique, where $\nu_1$ is scanned over the all spin resonances and the RF at $\nu_2$ is modulated on/off allowing lock-in detection of the relative changes in the ODMR signal $\eta$ \cite{10.1038/s41467-019-09429-x}. It is normalized such that $\eta = 1$ for $\nu_1 > 45 \, \mathrm{MHz}$, i.e., above the spin transition at the highest available frequency where no changes in the ODMR signal are expected.  

Figures~\ref{fig3}(b) and (c) show the ODMR signals $\eta_{\uparrow} (\nu_1)$ and $\eta_{\downarrow} (\nu_1)$ as a function of the scanned RF $\nu_1$ for the modulated RF being fixed to $\nu_2 = \nu^{(\uparrow)}$ and $\nu^{(\downarrow)}$ respectively. A 2D scan when the both $\nu_1$ and $\nu_2$ are varied is presented in the Supplemental materials.
The ODMR signal exhibits various resonances, which correspond to the different 
spin transitions labeled in Fig.~\ref{fig3}(a). The matrix elements of all these transitions are summmirized in table~\ref{tab:hyperfine}. 
A remarkable features observed from the comparison of the spectra in Fig.~\ref{fig3}(b) and (c) is that $\eta_{\uparrow}(\nu_1) \neq \eta_{\downarrow}(\nu_1)$ in the vicinity of the 
transitions (1) and (2), i.e., for $\nu_1 = \nu^{(\uparrow)}$ and $\nu^{(\downarrow)}$. These two transitions have the same matrix element and differ only by the nuclear spin in their initial states. Therefore, the mismatch of their amplitudes is a direct manifestation of the nuclear spin polarization.

For the quantitative description of our experimental data, we develop a phenomenological model of DNP.
In a nutshell, the optically generated quadrupole polarization $P_Q$ of the electron spin  in Eq.~(\ref{eq:quadrupole}) is converted into the nuclear spin polarization by applying RF $\nu_1$. To detect this polarization, a weak modulated RF $\nu_2$ is applied, which partially converts the nuclear spin dipole back to electron spin quadrupole polarization. The latter is measured optically as the ODMR signal variation $\eta$.

In particular, the tuning of the weak modulated RF to $\nu_2 = \nu^{(\uparrow)}$ leads to the ODMR variation 
\begin{align}
    \eta_\uparrow (\nu_1)= \left[ \frac{p_{-3/2,\uparrow}(\nu_1)-p_{s}(\nu_1)}{p_{-3/2,\uparrow}^{(0)}-p_{s}^{(0)}} \right]^2 .
\end{align}
Here,  $p_{i}$ denotes the probability to be in the state $|i\rangle$ and the superscript $(0)$ corresponds to the probabilities in the absence of RF fields. The quadratic dependence of the signal on the population difference $p_{+3/2,\downarrow}-p_{s}$ originates from the fact that it determines not only the spin transition rates, but also quantifies how these rates affect the detected electron spin quadruple polarization (a detailed description is presented in the Supplemental materials). The ODMR variation $\eta_\downarrow (\nu_1)$ is described by a similar equation. 


Using the fact that in the absence of the RF fields $p_{-3/2,\uparrow}^{(0)}-p_{s}^{(0)} = p_{+3/2,\downarrow}^{(0)}-p_{s}^{(0)} = P_Q/4$, we can express 
\begin{align}
p_{-3/2,\uparrow}(\nu_1) - p_{-3/2,\downarrow}(\nu_1) =  \frac{P_Q}4 \left[\sqrt{\eta_\uparrow (\nu_1)} - \sqrt{\eta_\downarrow(\nu_1)} \right]  .
\end{align}
The obtained difference of probabilities contributes to the total nuclear polarization degree
$ P_N = P_\uparrow - P_\downarrow $, where $P_\uparrow$ and $P_\downarrow$ are the probabilities to find the $\mathrm{V_{Si}}$ center in the states with the nuclear spin $\uparrow$ and $\downarrow$ respectively. To find $P_N$, information about population of the all spin sublevels is required. However, if a certain spin relaxation mechanism is assumed, all the populations can be expressed via $p_{-3/2,\uparrow} - p_{-3/2,\downarrow}$. Then we obtain 
\begin{align}\label{eq:PN}
      P_N(\nu_1) = \frac{\kappa P_Q}{2} \left[\sqrt{\eta_\uparrow (\nu_1)} - \sqrt{\eta_\downarrow(\nu_1)} \right] ,
\end{align}
where $\kappa \sim 1$ is a dimensionless parameter that quantifies the relaxation mechanism. Assuming spherically-symmetric electron spin relaxation and no nuclear spin relaxation, $\kappa = (6T_f + 5T_d - T_p)/(4T_f +5T_d+ T_p)$ where $T_p$, $T_d$ and $T_f$ are the relaxation rates for the electron spin dipole, quarupole, and octupole respectively~\cite{10.1002/pssb.201700258,10.1038/s41467-019-09429-x}. For simplicity, in what follows we assume $\kappa =1$, which corresponds, e.g., to $T_p=T_d=T_f$~\cite{10.1038/s41467-019-09429-x}.

\begin{figure}[t]
\includegraphics[width=.45\textwidth]{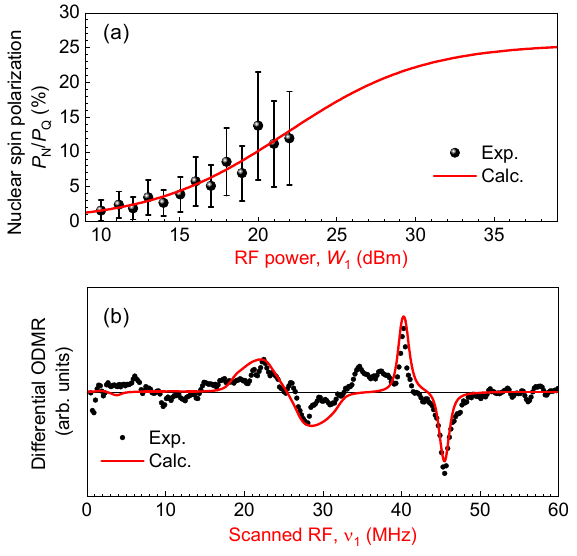}
\caption{DNP degree at $B = 64 \, \mathrm{\mu T}$. (a) Nuclear spin polarization $P_N$ relative to the quadrupole electron spin polarization $P_Q$ as a function of the scanned RF power $W_1$ at the fixed modulated RF power $W_2$. The circles represent values obtained from the experimental data using Eq.~(\ref{eq:PN}). The solid line is a fit to Eq.~(\ref{eq:Power}). (b) Differential ODMR spectrum between two modulated frequencies $\nu_2 = 45 \, \mathrm{MHz}$ and $\nu_2 = 40 \, \mathrm{MHz}$. The solid line is calculated using a model that includes inhomogeneous broadening as described in the text.} \label{fig4}
\end{figure}

Now, we use the aforementioned procedure to determine the nuclear polarization from the experimental ODMR signals in Fig.~\ref{fig3}(b,c). We fit the dips at $\nu_1 = 45 \, \mathrm{MHz}$ and $\nu_1 = 40 \, \mathrm{MHz}$ for $\nu_2= \nu^{(\uparrow)}= 45 \, \mathrm{MHz}$ [Fig.~\ref{fig3}(b)] and $\nu_2= \nu^{(\downarrow)}= 40 \, \mathrm{MHz}$ [Fig.~\ref{fig3}(c)] with a Gaussian function, as shown by the solid lines. Then, the dependence of the Gaussian amplitudes on the power of the scanned RF field is considered. To account for a possible small inhomogeneous broadening, the dependencies are rescaled in such a way that the depth of the dip at $\nu_1=\nu_2$ equals 0 at zero power and tends to 1 at large powers. The result is then plugged into Eq.~\eqref{eq:PN} to determine the DNP degree. The cyrcles in Fig.~\ref{fig4}(a) represent the obtained relative nuclear polarization $P_N/P_Q$ as a function of the RF power $W_1$ for $\nu_1 = 40 \, \mathrm{MHz}$. We fit the experimental points with a phenomenological dependence 
\begin{align}\label{eq:Power}
  P_N (W_1) = P_N^{\rm (max)}\frac{W_1}{W_0 + W_1 }  , 
\end{align}
where $P_N^{\rm (max)}$ is the maximal value of the nuclear spin polarization that can be achieved at large RF-pump powers.
$W_0$ is a constant that quantifies the relaxation rate of the population difference between the relevant singlet and doublet states, see Supplemental materials for details.
The best fit yields $P_N^{\rm (max)}/P_Q =0.26$.

Figure~\ref{fig4}(b) shows the DNP as a function of $\nu_1$, extracted according to Eq.~\eqref{eq:PN}. The cyrcles are obtained from the experimental spectra of Fig.~\ref{fig3}(b,c), while the solid line stems from the theoretically calculated spectra, presented in the Supplemental materials. Apart from the peaks at $\nu_1 = 40$\,MHz and 45\,MHz described above, the spectra feature also two peaks at $\nu_1 = 22$\,MHz and 28\,MHz. The latter peaks correspond to the transitions between the same doublet states with $S_zI_z = -3/4$ and the other singlet state $|s'\rangle$. The DNP peaks at these frequencies  are significantly broadened compared to that at $\nu_1 = 40$\,MHz and 45\,MHz. We attribute this inhomogeneous broadening to the spread of the $\mathrm{V_{Si}}$ parameters across the sample. In our theoretical calculation, we assume a certain dispersion of $D$ and $A$. The lock-in detection  selects the $\mathrm{V_{Si}}$ centers with a certain combination of $D$ and $A$, for which the scanned frequency matches  the corresponding transitions $\nu^{(\downarrow, \uparrow)} = | 2D-3A/2 \pm 3g\mu_B B/2 |$ (see Table~\ref{tab:hyperfine}). This eliminates the inhomogeneous broadening for the peaks at $\nu_1 = 40$\,MHz and 45\,MHz. However, the transitions at frequencies  $\nu_1 = 22$\,MHz and 28\,MHz are described by a different combination of  $D$ and $A$, i.e., $ | 2D+A/2 \pm 3g\mu_B B/2 |$. Therefore the inhomogeneous broadening of these peaks persists. Using this approach, we achieve a perfect agreement with our experimental data of Fig.~\ref{fig4}(b).

\section{Conclusion and outlook}

In summary, we demonstrate  dynamic $\mathrm{^{29}}$Si nuclear polarization in 6H-SiC at room temperature by applying a very weak magnetic field, which is in the range of the Earth's magnetic fields. The polarization protocol is based on the  optical pumping of the $\mathrm{V_{Si}}$ centers with half-integer electron spin $S = 3/2$ and asymmetric  hyperfine structure. Our theoretical model provides a very good agreement with the experimental data. 

Our DNP protocol relies on an efficient conversion of the quadrupole electron $\mathrm{V_{Si}}$ spin polarization to the nuclear spin polarization and we approach the conversion efficiency of 26\% in the limit of high RF power. For a high-fidelity electron spin initialization \cite{10.48550/arxiv.2401.04470, 10.48550/arxiv.2401.04465}, this conversion efficiency corresponds to an effective nuclear temperature of $\pm 50 \, \mathrm{n K}$. Here, the negative temperature means population inversion of the nuclear spin states. This value is by one order of magnitude lower than that reported for ultra-deep optical cooling of nuclear spins in GaAs quantum wells \cite{10.1038/s42005-021-00681-6} and two orders of magnitude lower than an effective nuclear temperature achieved in SiC using another DNP protocol  \cite{10.1103/physrevlett.114.247603}. 

The requirement of a weak magnetic field and no need for precise magnetic field alignment make our protocol extremely easy to use. Particularly, it is attractive  for the implementation of nuclear hyperpolarization imaging, where SiC nanostructures can be used either for the direct enhancement of the image contrast or for the transferring of the nuclear polarization to contrast agents. Our approach also suggests an alternative protocol for the initialization and coherent control of nuclear spins in SiC with long lived quantum memory, where magnetic field homogeneity is of crucial importance. Furthermore, due to suppression of inhomogeneous broadening for the certain Zeeman-split transitions, the electron-nuclear spin registers can be used for quantum sensing with enhanced sensitivity. 


\section*{Acknowledgments}
This work has been co-financed by tax funds based on the budget passed by the Saxon State Parliament.
A.V.P. acknowledges support from the Government of Spain under Severo Ochoa Grant CEX2019-000910-S [MCIN/AEI/10.13039/501100011033], Generalitat de Catalunya (CERCA program), Fundaci\'{o} Cellex, and Fundaci\'{o} Mir-Puig.


\bibliography{SiC_NucleiSpin_literature} 

\end{document}




\title{Supplemental materials for \\ Nuclear spin polarization in silicon carbide at room temperature in the Earth's magnetic field}

\author{A.~N.~Anisimov$^{1}$}
\email[E-mail:~]{a.anisimov@hzdr.de}
\author{A.~V.~Poshakinskiy$^{2}$}
\author{G.~V.~Astakhov$^{1}$}
\email[E-mail:~]{g.astakhov@hzdr.de}

\affiliation{$^1$Helmholtz-Zentrum Dresden-Rossendorf, Institute of Ion Beam Physics and Materials Research, 01328 Dresden, Germany  \\
$^2$ICFO-Institut de Ciencies Fotoniques, The Barcelona Institute of Science and Technology, 08860 Castelldefels, Barcelona, Spain
 }



\maketitle

\begin{figure}[t]
\includegraphics[width=.45\textwidth]{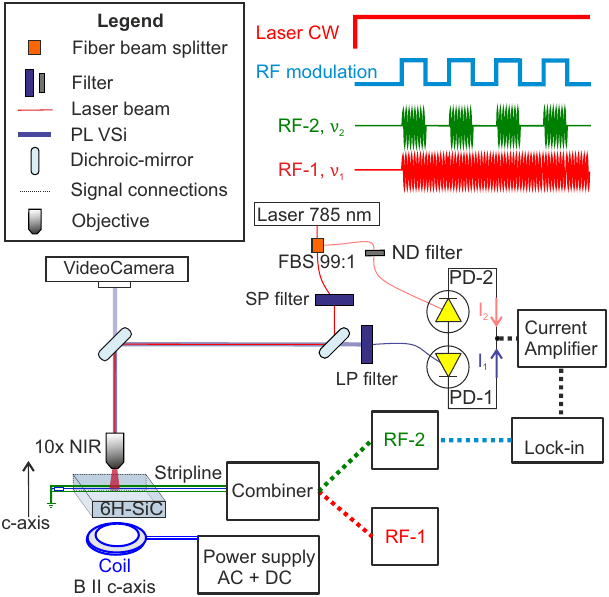}
\caption{Simplified diagram of the ODMR setup. Electronic components are represented by the rectangles. The legend for the optical components is provided in the top left corner. 
The RF signals at frequencies $\nu_1$ and $\nu_2$ from signal generators RF-1 and RF-2 (the dashed red and green lines, respectively) are combined and delivered to a stripline. 
The signal from the balanced photodetection circuit is shown with the black dashed line. The solid dark blue and red lines connected to PD-1 and PD-2 depict optical fibers. The currents generated by PD-1 and PD-2, shown with the dark blue and red arrows, are equal in magnitude without RF fields and directed oppositely. The differential current is converted to voltage and locked-in.  }   \label{SM_fig1}
\end{figure}

\begin{figure}[t]
\includegraphics[width=.45\textwidth]{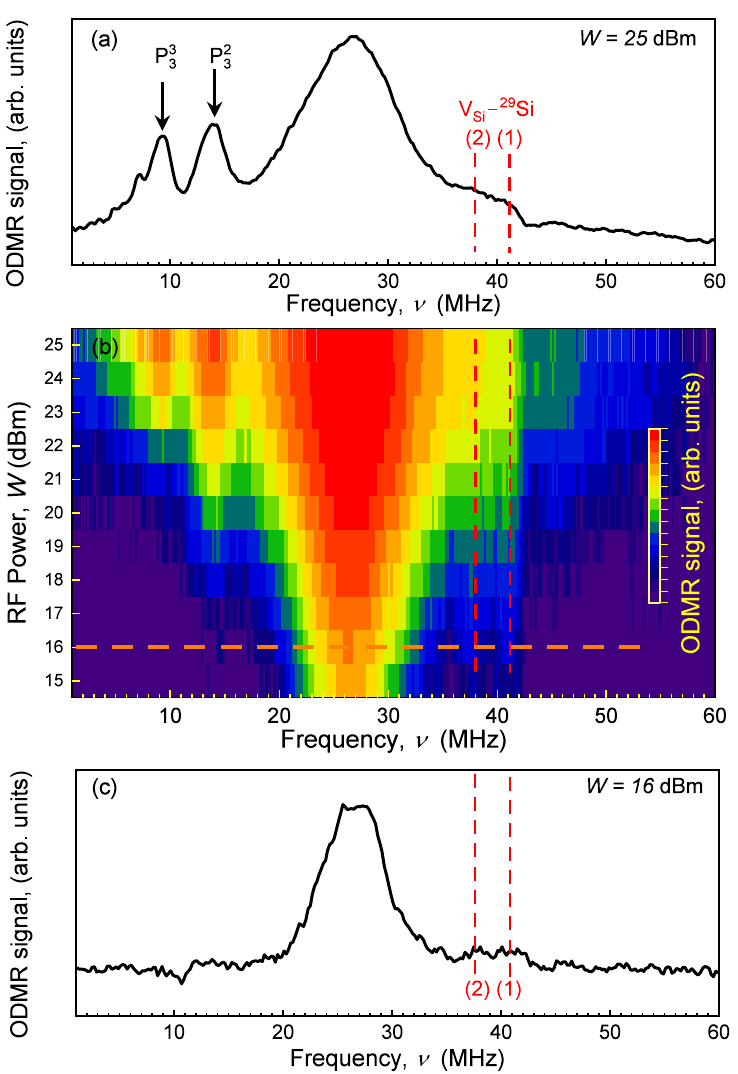}
\caption{RF-power dependence of the V3 $\mathrm{V_{Si}}$ ODMR spectrum in 6H-SiC at room temperature in an uncompensated environment magnetic field. (a) ODMR spectrum for $W=25 \, \mathrm{dBm}$ power. Vertical arrows indicate the positions of peaks corresponding to the two-photon $\nu=14 \, \mathrm{MHz}$ ($P_2^3$) and three-photon $\nu=9 \, \mathrm{MHz}$ ($P_3^3$) transitions. These peaks occur at pumping powers greater than $W= 16 \, \mathrm{dBm}$.  Red dashed lines on all graphs indicate the positions of transitions (1) and (2), corresponding to the hyperfine interaction of the silicon vacancy ($\mathrm{V_{Si}}$) with nuclei $\mathrm{^{29}}$Si. (b) A 2D plot showing the frequency dependence on pump power. The intensity of the ODMR signal is depicted in a logarithmic scale. The horizontal orange dashed line indicates the optimal power ($W=16 \, \mathrm{dBm}$) at which the ODMR signal is not broadened, and transitions related to nuclei $\mathrm{^{29}}$Si spins and two- and three-photon transitions have insignificant contribution. (c) ODMR spectrum for the optimal power of $W=16 \, \mathrm{dBm}$. Vertical dashed lines indicate the peak positions corresponding to the hyperfine interaction of the $\mathrm{V_{Si}}$ with $\mathrm{^{29}}$Si nuclei. 
} \label{SM_fig5}
\end{figure}

\begin{figure}[t]
\includegraphics[width=.45\textwidth]{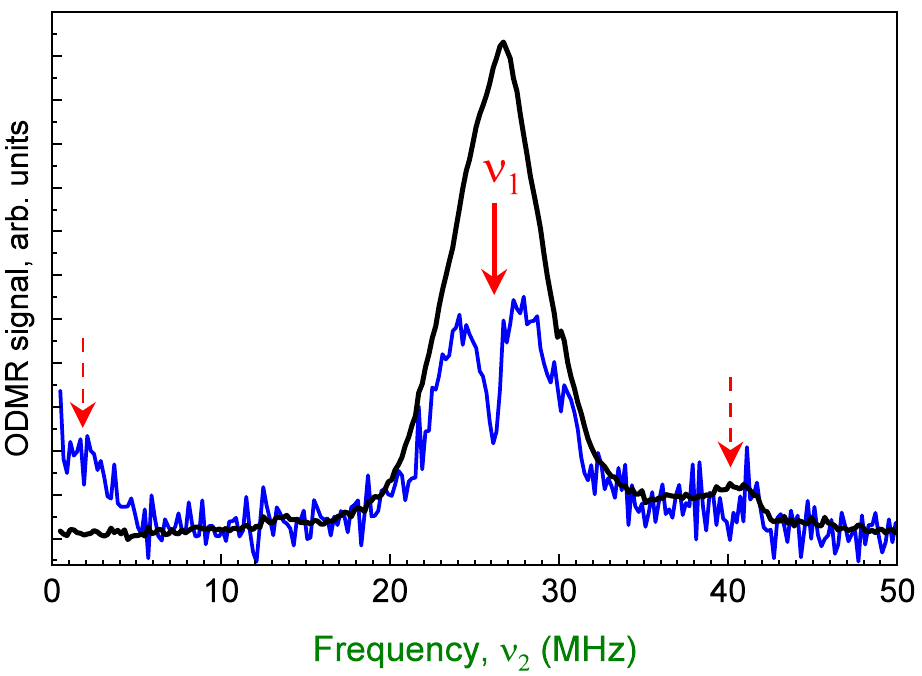}
\caption{Two-frequency room-temperature ODMR spectroscopy for the V3 $\mathrm{V_{Si}}$ center in 6H-SiC. The ODMR spectrum is shown without (black curve) and with (blue curve) the additional unmodulated RF field at $\nu_1 = 26 \, \mathrm{MHz}$. The pump and probe powers are $W=16 \, \mathrm{dBm}$. The red arrow shows the result of the applied RF field. }   \label{SM_fig2}
\end{figure}

\begin{figure}[t]
\includegraphics[width=.45\textwidth]{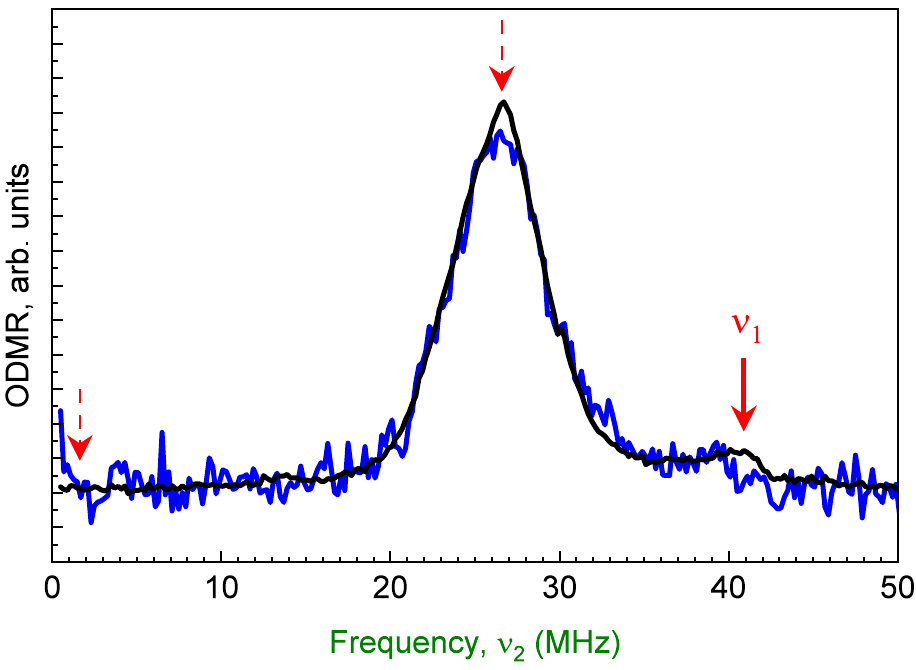}
\caption{Two-frequency room-temperature ODMR spectroscopy for the V3 $\mathrm{V_{Si}}$ center in 6H-SiC.The ODMR spectrum is shown without (black curve) and with (blue curve) the additional unmodulated RF field at  $\nu_1 = 41 \, \mathrm{MHz}$. The pump and probe powers are $W=16 \, \mathrm{dBm}$. The red arrow shows the result of the applied RF field.} \label{SM_fig3}
\end{figure}

\section{\label{sec:level1}Experimental setup}

Figure~\ref{SM_fig1} shows a schematic of the optically detected magnetic resonance (ODMR) spectrometer. Optical excitation is performed using a semiconductor laser with a wavelength $\lambda = 785 \, \mathrm{nm} $ (FPL785S-250) and an excitation output power of about $100 \, \mathrm{mW}$. The laser is split at a ratio of 99:1 using a fiber beam splitter (FBS), where $99\%$ is sent into the sample pump channel and $1\%$ to the photodiode $2$ (PD-2).  A fiber-coupled variable neutral density (ND) filter is used to adjust the laser power delivered to PD-2.   To clean the laser excitation, a $800\mathrm{nm}$  short-pass filter (FESH800) (SP filter) is placed in front of the dichroic mirror (DMSP850R) at the laser output. The photoluminescence (PL) is excited and collected  through an 10x objective (LMPLN10XIR from Olympus) and a second dichroic mirror. A long-pass filter is installed in the PL collection channel, filtering out the laser radiation before it enters the detection channel of the photodiode 1 (PD-1). PD-1 and PD-2 are connected in a balanced photodetection circuit. To monitor the measurement process, the spectrometer is equipped with a camera that captures a small portion of light from the sample. 

For radiofrequency (RF) pump-probe experiments, two  generators (RF-1 and RF-2) are used, one of which is modulated  at 2 kHz and synchronized with a lock-in detector. We use $100\%$ amplitude modulation for generator 2 and an RF power amplifier (ZX60-100VH+ and ZHL-10W-202-S+ from Mini Circuits). Both frequencies are combined using a combiner and  delivered to the focal area of the objective using a high-frequency stripline. The setup is equipped with XYZ mechanical stages for  the selection of the area for optical detection. 
Changes in the PL signal $\Delta$PL are detected using a balanced photodetection scheme and a current amplifier. It compensates optical noise created by the laser. 

An example of such spectra is shown in Figs~\ref{SM_fig2} and \ref{SM_fig3}. For the level anticrossing (LAC) spectroscopy \cite{10.1007/s00723-017-0938-1}, a bias magnetic field generated by a magnetic coil along the SiC c-axis is used. Additionally, a modulation coil connected to the lock-in detector is used for the amplitude modulation of the magnetic field. A power supply unit is employed to create both the constant (DC) and alternating (AC) components of the magnetic field. The top of Figure ~\ref{SM_fig1} schematically shows the control and pump signals. 

\section{\label{sec:level1} ODMR optimization}

A 2D map of the evolution of the V3 $\mathrm{V_{Si}}$ ODMR spectrum with increasing RF power is presented in Fig.~\ref{SM_fig5}(b). Figures~\ref{SM_fig5}(a) and (c) show the ODMR spectra at RF powers $W=25 \, \mathrm{dBm}$ and $16 \, \mathrm{dBm}$, respectively. 
By comparing these two spectra, 
a different number of the ODMR peaks is observed. This is because at $25 \, \mathrm{dBm}$ the system is over-pumped, and we observe two-photon transitions at $\nu=14 \, \mathrm{MHz}$ ($P_2^3$) and three-photon transitions at $\nu=9 \, \mathrm{MHz}$ ($P_3^3$) \cite{10.1103/PhysRevResearch.4.023022}. These transitions are indicated by the black arrows in Fig.~\ref{SM_fig5}(a). 
To optimize the RF power, we measured the dependence of the ODMR signal on RF power $W$ in the range from 15 to 25$\, \mathrm{dBm}$. This dependence is shown in Figure~\ref{SM_fig5}(b), where the horizontal orange dashed line indicates the optimal power $W=16 \, \mathrm{dBm}$. At this power, the contribution to the ODMR spectrum from silicon vacancies related to hyperfine interactions with $\mathrm{^{29}}$Si nuclei is much stronger than the contribution from two- and three-photon transitions \cite{10.1103/PhysRevResearch.4.023022}.  Moreover, the state degeneracy in the $\mathrm{V_{Si}}$ center associated with the antiparallel $\mathrm{^{29}}$Si nuclear orientations is lifted up in an external magnetic field, allowing for targeted addressing of the nuclear spin using a two-frequency technique. 

\section{\label{sec:level1} Two-frequency ODMR}

Figures~\ref{SM_fig2} and \ref{SM_fig3} compare single-frequency ODMR spectroscopy (black curve) and two-frequency ODMR spectroscopy (blue curve) for the V3 $\mathrm{V_{Si}}$ in 6H-SiC. The peak at  $| 2D |= 28 \, \mathrm{MHz}$ is associated with the zero-field splitting (ZFS) between the Kramers doublets $m_{S}=\pm1/2$ and $m_{S} = \pm3/2$. This ODMR spectrum is recorded using lock-in detection of the PL change under amplitude modulation of the RF field, which is scanned from $\nu_2 = 1 \, \mathrm{MHz}$ to $\nu_2 = 50 \, \mathrm{MHz}$ (black curve in Fig.~\ref{SM_fig2}). When the second RF field is applied at a fixed frequency $\nu_1 = 26\, \mathrm{MHz}$, the ODMR spectrum changes (blue curve in Fig.~\ref{SM_fig2}). 
The solid arrow shows a dip at $\nu_2 = \nu_1 = 26\, \mathrm{MHz}$ directly related to the ZFS. The dashed arrows represent transitions at different frequencies, which occur due to changes in the occupation of spin states caused by the RF pump at $\nu_1 = 26\, \mathrm{MHz}$. The low-frequency transition couples the electron spin $m_{S}=+1/2$ and $m_{S} = -1/2$ states \cite{10.1038/s41467-019-09429-x}. The high-frequency transition with a dip at $41 \, \mathrm{MHz}$ is associated with the hyperfine interaction with the $^{29}$Si  nuclear spin.  

We explore the potential of two-frequency ODMR for targeted probing of the electron-nuclear spin system by tuning the frequency of the pump RF field to $\nu_1 = \, 41 \mathrm{MHz}$ (blue curve in Fig.~\ref{SM_fig3}). The dips at the frequencies indicated by the dashed arrows are significantly smaller than in Fig.~\ref{SM_fig2}. 
This is because these dips are associated only with the electron-nuclear spin system, and their contribution to the overall signal, which includes not only the non-zero $^{29}$Si but also zero $^{28}$Si nuclear spin, is relatively weak. 

\begin{figure}[t]
\includegraphics[width=.45\textwidth]{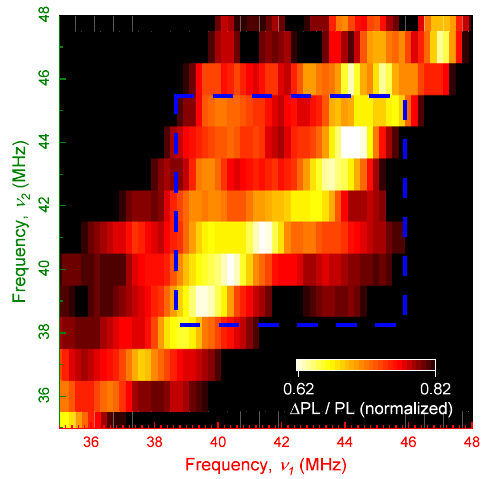}
\caption{A 2D color map of two-frequency ODMR in the spectral window corresponding to the $^{29}$Si nuclear transitions in the V3 $\mathrm{V_{Si}}$ center. The power of the RF fields is the same $16\, \mathrm{dBm}$. The square restricted by the blue dashed line indicates the frequency region where the dynamic nuclear polarization is detected. The strongest ODMR signal is observed when the frequencies coincide, which corresponds to the diagonal of the square.  } \label{SM_fig4}
\end{figure}

In order to detect only the electron spins coupled to the $^{29}$Si  nuclei, we swap the frequencies $\nu_1$ and $\nu_2$. We then scan the frequency $\nu_1$ of the first RF field while amplitude-modulate the second RF field, which is fixed at a frequency $\nu_2$ related to the spin transition associated with the $^{29}$Si nuclear spin. This technique is used for the dynamic nuclear polarization (DNP). The results of the DNP are presented in the main text in Fig.~3, where we additionally use the Earth's magnetic field to split the spectral lines associated with different nuclear spin orientations. To determine the region where DNP occurs, we perform 2D scanning of the frequencies $\nu_1$ and $\nu_2$ in the range of $35 - 48$~MHz (Fig.~\ref{SM_fig4}). The suitable region for the DNP is indicated by the blue dashed square. The strongest ODMR signal is observed when the frequencies coincide, corresponding to the diagonal of the square. The brightest areas in the 2D map are associated with the frequencies 40 and 45~MHz, which are used for the DNP, as described in the main text.

\section{Theoretical model}

The state of an ensemble of spin-3/2 centers that coupled to adjacent spin-1/2 nuclei is described by the $8 \times 8$ density matrix $\rho$. Its evolution  is governed by the master equation
\begin{align}\label{eq:Lind}
   \frac{d\rho}{dt} = \frac{\rmi}{\hbar} [\rho, H] + R[\rho] + P[\rho], 
\end{align}
where $H$ is the Hamiltonian -- see Eq.~(1) in the main text -- $R$ and $L$ are the Lindbladians, that account for the optical generation of the quadrupolar electron spin polarization and spin relaxation respectively. In the simplest model that we use here, the relaxation times of all electron spin multipoles are assumed equal. Then,
\begin{align}
    R[\rho] = \Gamma_S \left( \frac{I_S}{4}\, {\rm Tr}_S \rho -\rho\right),
\end{align}
where $\Gamma_S=1/T_S$ is the relaxation rate, ${\rm Tr}_S$ indicates the trace in the subspace of the electron  spin, and $I_S$ is the identity matrix in this subspace.
A more general case where electron spin dipole, quadrupole and octupole have different relaxation times was introduced in Ref.~\cite{10.1038/s41467-019-09429-x}. The relaxation of the nuclear spin can be described in a similar way. However, we suppose that the nuclear spin relaxation time is much longer than that of the electrons, $T_N \gg T_s$, so the former can be neglected. The optcial generation of the quadrupolar electron spin polarization is described by
\begin{align}
    P[\rho] = \Gamma_P \left\{ \frac{1}{4}\left[I_S + \eta \left( S_z^2-\frac54 I_S \right) \right] {\rm Tr}_S \rho -\rho\right\} ,
\end{align}
where $\Gamma_P$ is the optical pump rate and $|\eta| \leq 1$ denotes the efficiency of the electron spin quadrupole generation.
The electron spin quadrupole polarization also determines the spin-dependent contribution to the PL intensity of the ensemble 
\begin{align}
    \Delta{\rm PL} \propto {\rm Tr} \left[\left( S_z^2-\frac54 I_S \right) \rho \right] .
\end{align}

To calculate ODMR spectrum, we solve master equation~\eqref{eq:Lind} in the presence of the two RF magnetic fields 
$B_1 \e^{-\rmi \omega_1 t} + B_2 \e^{-\rmi \omega_2 t} + {\rm c.c.}$. To this end, we use for the density matrix the expression
\begin{align}
    \rho (t) = \rho_0 + \rho_1 \e^{-\rmi \omega_1 t} + \rho_2 \e^{-\rmi \omega_2 t} + \rho_1^\dag \e^{\rmi \omega_1 t} + \rho_2^\dag \e^{\rmi \omega_2 t} ,
\end{align}
where the higher-order harmonics are neglected. 

\begin{figure}[tb!]
\includegraphics[width=.45\textwidth]{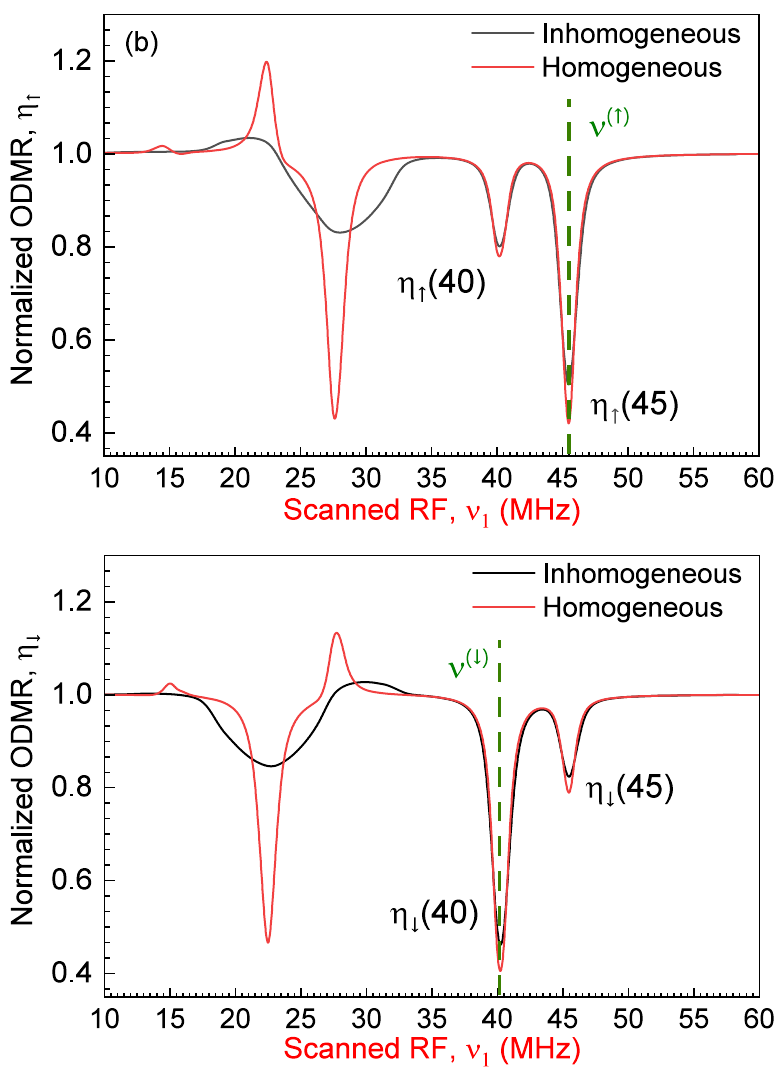}
\caption{ODMR signals $\eta_\uparrow(\nu_1)$ and $\eta_\downarrow(\nu_1)$ calculated using the developped theoretical model. The red curves correspond to the absence of the inhomogeneous broadening. For the black curves, the fluctuations of the zero field splitting $D$ and hyperfine interaction constant $A$ with the variances $\text{var}(D/2\pi) = \text{var}(A/2\pi) = 3\,\text{MHz}^2$ were taken into account. Other parameters are the same for the two cases and read  $D/h=-14\,$MHz, $A/h=8.84\,$MHz, $B_z = 63.5\,\mu$T, $\Gamma_S/(2\pi) = 0.3\,$MHz, $\Gamma_P/(2\pi) = 0.001\,$MHz, $\eta=1$, $B_1 = 11\,\mu$T, $B_2 = 6\,\mu$T and $\bm B_1, \bm B_2 \parallel x$.\\
} \label{fig:Seta}
\end{figure}

Figure~\ref{fig:Seta} shows the normalized ODMR signal calculated according to the described model. The red curves correspond to the absence of the inhomogeneous broadening. For the black curves, the signal (before normalization) was averaged over an ensemble of centers with different values  of the zero field splitting $D$ and hyperfine interaction constant $A$.

\section{Rate equations}

Here we suppose that the density matrix is diagonal in the basis of the eigen states of the static part of the Hamiltonian. The two RF fields are supposed to be resonant with certain transitions between the levels $k_i^{(1,2)}$ and $k_f^{(1,2)}$: $\hbar\omega_{1,2} = |E_{k_i^{(1,2)}} - E_{k_f^{(1,2)}}|$.
The dynamics of the diagonal elements $p_i$, $i=1,2,\ldots 8$, is governed by the rate equation
\begin{align}\label{eq:R}
    \frac{dp_i}{dt} = -\sum_j [\Gamma_{ij} + W_1 t^{(1)}_{i}t^{(1)}_{j} + W_2 t^{(2)}_{i}t^{(2)}_{j}]p_j + \Gamma_P q_i 
\end{align}
where $\Gamma$ is the symmetric matrix describing the spin relaxation, $W_{1,2}$ is the rate of the spin transitions induced by the two RF fields, $t^{(1,2)}_{j} = \delta_{j,k_i^{(1,2)}}- \delta_{j,k_f^{(1,2)}} $, and $q_i = (S_z^2)_{ii}-5/4$ is the optically generated quadrupole. 
Note that due to conservation of $\sum_i p_i =1$, the number of independent variables is in fact reduced to 7.

The steady-state solution of the rate equation~\eqref{eq:R} can be presented in a matrix format as
\begin{align}\label{eq:dp}
    \bm p = \Gamma_P (\bm \Gamma + W_1 \, \bm t^{(1)} \otimes \bm t^{(1)} + W_2\, \bm t^{(2)} \otimes \bm t^{(2)})^{-1} \bm q .
\end{align}
Now we suppose that $W_2 \ll \Gamma_S$, and find the variation of the populations corresponding corresponding to on/off modulation of $W_2$,
\begin{align}
    \Delta \bm p = -\Gamma_P W_2\,  \bm T^{(1)} \bm t^{(2)} (\bm t^{(2)} \cdot \bm T^{(1)} \bm q ) .
\end{align}
where we introduced $\bm T^{(1)} = (\bm \Gamma + W_1 \, \bm t^{(1)} \otimes \bm t^{(1)} )^{-1}$. Note that 
\begin{align}\label{eq:Sp1}
    \bm p^{(1)} \equiv \bm T^{(1)} \bm q = \bm\Gamma^{-1}\bm q - \bm\Gamma^{-1}\bm t\, \frac{W_1 \bm t^{(1)} \cdot \bm \Gamma^{-1} \bm q}{1+ W_1 \bm t^{(1)} \cdot \bm \Gamma^{-1} \bm t^{(1)}}
\end{align}
are the equilibrium populations in the presence of the first RF field when the second RF field is turned off.
In accord with Eq.~\eqref{eq:Sp1}, the variations of the populations with the intensity of the RF field $W_1$ should follow the dependence 
\begin{align}
    \Delta p^{(1)} (W_1) \equiv p^{(1)} (W_1)- p^{(1)} (0) \propto \frac{W_1}{1 + W_1/W_0}
\end{align}
with the same $W_0=1/(\bm t^{(1)} \cdot \bm \Gamma^{-1} \bm t^{(1)})$ for all levels. The nuclear polarization degree, being determined by some combination of the populations, follows the same phenomenological dependence.

The PL variation corresponding to the population modulation induced by the second RF field Eq.~\eqref{eq:dp} is given by $\bm q \cdot \Delta \bm p$ and reads
\begin{align}
    \Delta{\rm PL} \propto -\Gamma_P W_2\,  (\bm q \cdot \bm T^{(1)} \bm t^{(2)}) (\bm t^{(2)} \cdot \bm T^{(1)} \bm q ).
\end{align}
Using the fact that the matrix $T^{(1)}_{ij}$ is symmetric, this can be simplified to
\begin{align}
    \Delta{\rm PL} \propto -\Gamma_P W_2\,  \left[ p^{(1)}_{k_i^{(2)}} - p^{(1)}_{k_f^{(2)}} \right]^2.
\end{align}
This signal is then normalized by diving it by its value for detuned first RF, i.e., for $W_1 =0$. This gives
\begin{align}
    \eta =  \left( \frac{p^{(1)}_{k_i^{(2)}} - p^{(1)}_{k_f^{(2)}}}{p^{(0)}_{k_i^{(2)}} - p^{(0)}_{k_f^{(2)}}} \right)^2.
\end{align}
where $\bm p^{(0)} = \bm \Gamma^{-1} \bm q $ are the level populations in the absence of RF fields.


\bibliography{SiC_NucleiSpin_literature} 